\documentclass[aps,prb,twocolumn,superscriptaddress]{revtex4}
\usepackage{epsfig}
\usepackage{graphicx}
\usepackage{dcolumn}
\usepackage{bm}

\begin{document}
\title{Structure of nonuniform hard sphere fluids from shifted linear truncations
of functional expansions}
\author{Yng-Gwei Chen}
\affiliation{Institute for Physical Science and Technology,}
\affiliation{and Department of Physics, University 
of Maryland, College Park, MD 20742}
\author{John D. Weeks}
\affiliation{Institute for Physical Science and Technology,}
\affiliation{and Department of Chemistry and Biochemistry, University 
of Maryland, College Park, MD 20742}
\date{\today }

\begin{abstract}
Percus showed that approximate theories for the structure of nonuniform hard
sphere fluids can be generated by linear truncations of functional
expansions of the nonuniform density $\rho ({\bf r})$ about that of an
appropriately chosen uniform system. We consider the most general such
truncation, which we refer to as the shifted linear response (SLR) equation,
where the density response $\rho ({\bf r})$ to an external field $\phi ({\bf r})$
is expanded to linear order at each ${\bf r}$ about a different uniform system with
a locally shifted chemical potential. Special cases include the Percus-Yevick (PY)
approximation for nonuniform fluids, with no shift of the chemical potential,
and the hydrostatic linear response (HLR)\ equation, where the chemical
potential is shifted by the local value of $\phi ({\bf r})$.
The HLR equation gives exact results for very slowly varying $\phi ({\bf r%
})$ and reduces to the PY approximation for hard core $\phi ({\bf r}),$
where generally accurate results are found. We show that a truncated
expansion about the bulk density (the PY approximation) also gives exact
results for localized fields that are nonzero only in a ``tiny'' region whose
volume $V^{\phi }$ can accommodate at most one particle. The SLR equation can also
exactly describe a limit where the fluid is confined by hard walls to a very narrow
slit. This limit can be related to the localized field limit by a simple shift of
the chemical potential, leading to an expansion about the ideal gas. We then
try to develop a systematic way of choosing an optimal local shift in the
SLR equation for general $\phi ({\bf r})$ by requiring that the predicted
$\rho ({\bf r})$ is insensitive to small variations about the appropriate
local shift, a property that the exact expansion to all orders would obey.
The resulting insensitivity criterion (IC) gives a theory that reduces to the HLR
equation for slowly varying $\phi ({\bf r}),$ and is much more accurate
than HLR both for very narrow slits, where the IC agrees with exact results, and
for fields confined to ``tiny'' regions, where the IC gives very accurate (but not exact)
results. However, the IC is significantly less accurate than the PY and HLR
equations for single hard core fields. Only a small change in the predicted
reference density is needed to correct this remaining limit.
\end{abstract}

\maketitle

\section{Introduction}

In this paper we discuss approximate methods for determining the structure
and thermodynamics of nonuniform hard sphere fluids in a general external
field $\phi ({\bf r}).$ The field can directly describe the effects of fixed
solute particles, confining walls, and other sources of nonuniformity on the
hard sphere fluid \cite{frisch64,hansenmac,rowlinson89} and the model serves
as a reference system through which properties of nonuniform simple liquids
can be determined by density functional theory \cite{evans92} or molecular
field theory.\cite{weeks02}

We start from a general theoretical perspective first described in a classic
article by Percus,\cite{percus64} and reviewed below in section \ref
{sec:Exact expansions}. Percus suggested that approximate theories for the
density response to $\phi ({\bf r})$ could be generated by truncations of
formally exact functional Taylor series expansions of the nonuniform density 
$\rho ({\bf r})$ about that of an appropriately chosen uniform system. In
particular he argued that the density response to a hard core $\phi ({\bf r}%
) $ could be accurately described by a simple {\em linear truncation} of an
expansion about the {\em bulk} density. This yields the {\em Percus-Yevick
(PY) approximation} for nonuniform fluids, \cite{singletPY} discussed in
section \ref{sec:Confined fields}, and indeed this gives rather accurate
results for hard core solutes of varying diameters.\cite
{henderson76,henderson92,katsov01,chen04} The linear truncation is the key
to developing a practically useful theory, since higher order terms in the
expansions are very complicated. Unfortunately, the results of the PY
approximation deteriorate very quickly for more extended fields, especially
ones with attractive regions, where the linear extrapolation from the bulk
becomes very inaccurate.

In sections \ref{sec:Exact shifting} and \ref{sec:SLR} we exploit an exact
property of the grand canonical ensemble, where both the chemical potential $%
\mu ^{B}$ and external field $\phi ({\bf r})$ can be shifted by a constant
without changing any properties. \cite{evans92} This allows us to consider a
more flexible expansion where the density response $\rho ({\bf r})$ at {\em %
each} ${\bf r}$ is expanded to linear order about a {\em different} uniform
system with a locally shifted chemical potential.\cite
{weeks02,katsov01,chen04} We can use the additional freedom permitted by
this local shift to minimize errors arising from the linear truncation of
the expansion. Indeed, we show below that there are several different limits
where {\em exact} results can be obtained from a linearly truncated
expansion by using an appropriate shift.

When the field $\phi ({\bf r})$ is very slowly varying, it is natural to
expand about a uniform system where the chemical potential is shifted by the
local value of the field. This leads to the {\em hydrostatic linear response}
(HLR) equation, which we derived earlier using physically motived arguments.%
\cite{weeks02,katsov01,chen04} The HLR equation gives exact results for very
slowly varying $\phi ({\bf r})$ and turns out also to give the same
predictions as the PY approximation for hard core $\phi ({\bf r}).$ Thus it
can be successfully used for a much wider class of fields (both repulsive
and attractive). However, the specific choice for the shift made in the HLR
equation is based on the local value of the field, and this can give poor
results in certain special cases, particularly when there are rapid
variations in $\phi ({\bf r})$ in a confined region of space much smaller
than the correlation length of the bulk fluid.

But we show in section \ref{sec:SLR} that in two such cases a linearly
truncated expansion can again give exact results with different choices of
the locally shifted system. In particular we show that exact results can be
found by expanding about the bulk density (i.e., the PY approximation) for
external fields $\phi ({\bf r})$ that are nonzero only in a tiny region of
space whose volume $V^{\phi }$ is so small that the center of at most one
hard sphere can be accommodated within it. The fact that the PY
approximation can give exact results for such confined or ``tiny fields''
does not seem to be known in the literature. This limit is of more than
academic interest since some recent advances in density functional theory
have resulted from requiring that it be satisfied.\cite{tarazona00}
Moreover, we show how expansions used in the tiny field theory can be easily
modified to exactly describe a seemingly very different problem, a narrow
slit, where the potential is infinite except in a narrow rectangular slit of
width $L_{s}.$ In the limit as $L_{s}\rightarrow 0,$ we show that a
truncated expansion about the ideal gas can give exact results, \cite
{hendersonslit} whereas the PY (and HLR) equation gives very poor results in
this limit.\cite{henderson92}

The fact that different shifted linear truncations can exactly satisfy all
these distinct limits suggests that a generally useful theory might arise
from finding a local shift that in some sense minimizes the errors arising
from the linear truncation. In section \ref{sec:new criterion} we present
our first attempt along these lines, where the local shift is chosen
self-consistently so that the predicted density response $\rho ({\bf r})$ is
insensitive to small variations about the proper local choice, a property
that an exact expansion to all orders would satisfy. This ``insensitivity
criterion'' (IC) effectively generates an expansion about a local uniform
system whose density can be interpreted as a smoothed weighted average of
the full density $\rho ({\bf r}),$ reminiscent of results in certain
versions of weighted density functional theory.\cite{evans92,meister85}

Results of the various theories for the density response to several
different fields are presented in section \ref{sec:results}. In general the
IC performs very well: it reduces to the exact HLR equation for slowly
varying fields, gives exact results in the narrow slit limit, and generally
accurate (though not exact) results for tiny fields. However, the IC
approach is significantly less accurate than the PY or HLR equations for a
single hard core field, where it substantially overestimates the contact
density at high density. Since this is one limit where the PY and HLR
equations do give satisfactory results, this may not be of much practical
importance. But it would be better to have a general
approach that remains accurate for these cases as well. Suggestions for further
work along these lines and final remarks are given in section \ref{sec:Final
remarks}.

\section{Exact expansions for the density response to an external field}

\label{sec:Exact expansions}The potential distribution theorem of Widom \cite
{widom82} relates the density $\rho ({\bf r})$ of a single component hard
sphere fluid in an external field $\phi ({\bf r})$ to the probability $P(%
{\bf r};[\rho ])$ of inserting a hard sphere test particle at the position $%
{\bf r}$: 
\begin{equation}
\rho ({\bf r})=\Lambda ^{-3}e^{\beta [\mu ^{B}-\,\phi ({\bf r})]}P({\bf r}%
;[\rho ]).  \label{eq:sumrule}
\end{equation}
The direct influence of the external field on the test particle is excluded
in determining $P({\bf r};[\rho ]),$ which depends only on the
intermolecular interaction energy between the test particle and the hard
sphere fluid. Equivalently, $P({\bf r};[\rho ])$ is the probability that a
cavity whose radius is equal or greater than the diameter $d$ of the hard
sphere particles exists at the particular point ${\bf r}$, since only then
can the test particle be inserted. Here $\mu ^{B}$ is the chemical
potential, $\Lambda $ the thermal de Broglie wavelength, and $[\rho ]$
specifies the functional dependence on the density profile. $P({\bf r};[\rho
])$ can be formally reexpressed \cite{hansenmac,rowlinson89} in terms of the 
{\em one-body direct correlation function\ }$c^{(1)}({\bf r};[\rho ])$ as 
\begin{equation}
P({\bf r};[\rho ])=e^{c^{(1)}({\bf r};[\rho ])}.  \label{eq:prob}
\end{equation}

By expanding $P({\bf r}_{1};[\rho ])$ at a given ${\bf r}_{1}$ in a
functional Taylor series about a {\em uniform} fluid at some density $\tilde{%
\rho}$, we arrive at a formal expansion for the nonuniform density directly
related to an expansion suggested by Percus: \cite{percus64,percusnote} 
\begin{eqnarray}
\rho ({\bf r}_{1}) &=&\Lambda ^{-3}e^{\beta [\mu ^{B}-\phi ({\bf r}%
_{1})]+c^{(1)}(\tilde{\rho})}  \nonumber \\
&&\times \{1+\int d{\bf r}_{2}\,c^{(2)}(r_{12};\tilde{\rho})(\rho ({\bf r}%
_{2})-\tilde{\rho}) \nonumber \\
&&+Q({\bf r}_{1})+...\},
\label{eq:sumrunexpand}
\end{eqnarray}
where the quadratic term in the expansion is
\begin{eqnarray}
Q({\bf r}_1) &\equiv &\int d{\bf r}_{2}\int d{\bf r}%
_{3}\,\left\{ c^{(2)}(r_{12};\tilde{\rho})c^{(2)}(r_{13};\tilde{\rho}%
)\right. ~~~  \nonumber \\
&&\left. +c^{(3)}({\bf r}_{1},{\bf r}_{2},{\bf r}_{3};\tilde{\rho})\right\}
(\rho ({\bf r}_{2})-\tilde{\rho})(\rho ({\bf r}_{3})-\tilde{\rho})\,\,\,\,
\label{eq:Qexpand}
\end{eqnarray}
The $c^{(n)}({\bf r}_{1},{\bf r}_{2},...,{\bf r}_{n};[\rho ])$ are defined
by successive functional derivatives of $c^{(1)}({\bf r};[\rho ])$ with
respect to the singlet density, i.e., 
\begin{equation}
c^{(n)}({\bf r}_{1},{\bf r}_{2},...,{\bf r}_{n};[\rho ])\equiv \frac{\delta
c^{(n-1)}({\bf r}_{1},{\bf r}_{2},...,{\bf r}_{n-1};[\rho ])}{\delta \rho (%
{\bf r}_{n})}.  \label{eq:Cndiff}
\end{equation}

Since we expand about a uniform fluid state, $c^{(2)}({\bf r}_{1},{\bf r}%
_{2};[\rho ])=c^{(2)}(r_{12};\tilde{\rho}),\,$where $r_{12}=|{\bf r}_{2}-%
{\bf r}_{1}|$, due to translational invariance. Although the density $\rho (%
{\bf r})$ can have discontinuities caused by the discontinuities in the
external field $\phi ({\bf r})$, representing, e.g., a hard wall or a
spherical cavity, $P({\bf r};[\rho ])$ is always continuous and smooth. \cite
{rowlinson89} The expansion in eq \ref{eq:sumrunexpand} is designed to take
advantage of the smoothness of $P({\bf r};[\rho ])$. The hope is that with
proper choice of $\tilde{\rho}$ the expansion can be truncated at some low
order and a relatively simple theory for $\rho ({\bf r})$ will result.

Percus considered many other expansions as well, some of which might seem
even more promising.\cite{percus64} For example, by directly expanding $%
c^{(1)}({\bf r}_{1};[\rho ])$ in eq \ref{eq:prob} in a Taylor series, we are
guaranteed that the resulting approximation for the density after a
truncation is always nonnegative, an exact and desirable property not always
produced by truncations of eq \ref{eq:sumrunexpand}. However as discussed in
sections \ref{sec:Confined fields} and \ref{sec:SLR} there is a well defined
limit where the expansion in eq \ref{eq:sumrunexpand} truncates {\em exactly}%
, in contrast to the corresponding expansion for $c^{(1)}$. Moreover we will
show that both the HLR and the PY equations can be derived from eq \ref
{eq:sumrunexpand}. This suggests that it offers a versatile starting point
for further research.

\section{Confined fields and the PY approximation}

\label{sec:Confined fields}One limit where the expansion in eq \ref
{eq:sumrunexpand} is particularly useful is when the external field $\phi $
is nonzero only a region much smaller than the correlation length of the
fluid. The analyticity of $P({\bf r};[\rho ])$ then ensures that its values
in the tiny region where $\phi $ is nonzero can be accurately determined by
making use of a {\em low order extrapolation }of its values {\em outside},
i.e., where $\phi $ vanishes. For such locally confined fields, it seems
clear that the expansion in eq \ref{eq:sumrunexpand} should be about $\tilde{%
\rho}=\rho ^{B}\equiv \rho (\mu ^{B})$, where $\rho (\mu )$ gives the
density of the uniform system with $\phi =0$ as a function of the chemical
potential $\mu $. In terms of the quantities appearing in eqs \ref
{eq:sumrule} and \ref{eq:prob} this can also be written as 
\begin{equation}
\rho ^{B}=\Lambda ^{-3}e^{\beta \mu ^{B}+c^{(1)}(\rho ^{B})}.
\label{eq:bulksum}
\end{equation}
It seems plausible that expansion only to {\em linear order} in eq \ref
{eq:sumrunexpand} could then give an accurate description of the fluid's
density response to a very confined field: 
\begin{eqnarray}
\rho ({\bf r}_{1}) &=&\Lambda ^{-3}e^{\beta [\mu ^{B}-\phi ({\bf r}%
_{1})]+c^{(1)}(\rho ^{B})}  \nonumber \\
&&\times \left[ 1+\int d{\bf r}_{2}\,c^{(2)}(r_{12};\rho ^{B})(\rho ({\bf r}%
_{2})-\rho ^{B})\right] .  \label{eq:PY0}
\end{eqnarray}
This qualitative argument will be made more precise in section \ref{sec:SLR}.
Using eq \ref{eq:bulksum}, this equation can be rewritten as 
\begin{equation}
\rho ({\bf r}_{1})=\rho ^{B}e^{-\beta \phi ({\bf r}_{1})}\left[ 1+\int d{\bf %
r}_{2}\,c^{(2)}(r_{12};\rho ^{B})(\rho ({\bf r}_{2})-\rho ^{B})\right] .
\label{eq:PY}
\end{equation}
If the direct correlation function $c^{(2)}(r_{12};\rho ^{B})$ of the
uniform bulk fluid is known, eq \ref{eq:PY} can be solved for the density $%
\rho ({\bf r})$ induced by the external field $\phi ({\bf r})$.

Equation \ref{eq:PY} is the {\em PY approximation} for nonuniform fluids.%
\cite{percus64} We argue below that it gives exact results for {\em any}
sufficiently localized $\phi ({\bf r}),$ as suggested by the extrapolation
argument. Moreover, the PY approximation is known to give reasonably
accurate results for the density response to {}larger spherical cavities
(including the hard wall limit), where $\phi ({\bf r})$ is infinitely
repulsive inside a spherical region of radius $R,$ but zero elsewhere. For
such potentials the exact result $\rho ({\bf r})=0$ is trivially obtained
from eq \ref{eq:PY} in regions where $\phi ({\bf r})$ is infinite because of
the Boltzmann factor $e^{-\beta \phi ({\bf r})}.$

Despite this success, the linear extrapolation using the bulk fluid in eq 
\ref{eq:PY} would be expected to give poor results for external fields that
remain finite and vary over extended regions, especially in negative regions
of the field where errors in the truncated series can be greatly magnified
by the Boltzmann factor $e^{-\beta \phi ({\bf r})}$. These are limits where
the PY approximation is known to be very inaccurate.\cite{henderson92}

\section{Exact shifting property of the grand ensemble}

\label{sec:Exact shifting}When the linear truncation is inaccurate, it may
seem difficult to make further progress, since the higher order terms in eq 
\ref{eq:sumrunexpand} are too complicated to use in practical calculations.
However, as noted by Percus, \cite{percus64} one does {\em not} have to
expand eq \ref{eq:sumrunexpand} about the bulk density defined by eq \ref
{eq:bulksum}, nor does one have to expand about the same bulk state for each 
${\bf r}$ value of $\rho ({\bf r})$. We can use this additional flexibility
to greatly extend the accuracy of different linear truncations of eq \ref
{eq:sumrunexpand}.

As is well known, it is the combination $\mu ^{B}-\phi ({\bf r})$ that
determines the density profile in the grand canonical ensemble, and not $\mu
^{B}$ and $\phi ({\bf r})$ separately.\cite{evans92} When both the chemical
potential and the external field are shifted by the same constant, the
system's properties should thus remain unchanged. This exact {\em shifting
property} \cite{katsov01} of the grand ensemble will play a key role in what
follows.

Using this shifting property in eq \ref{eq:sumrunexpand}, we consider a
general shifted chemical potential 
\begin{equation}
\tilde{\mu}^{{\bf r}_{1}}=\mu ^{B}-a^{{\bf r}_{1}}  \label{eq:murr}
\end{equation}
and a shifted external field 
\begin{equation}
\tilde{\phi}^{{\bf r}_{1}}({\bf r})=\phi ({\bf r})-a^{{\bf r}_{1}},
\label{eq:phirr}
\end{equation}
both of which are shifted from the original $\mu ^{B}$ and $\phi ({\bf r})$
by a constant $a^{{\bf r}_{1}}$ that in principle can depend parametrically
on the point ${\bf r}_{1}$ about which the expansion is carried out, as
indicated by the superscript ${\bf r}_{1}$. The
shifted chemical potential $\tilde{\mu}^{{\bf r}_{1}}$ defines at each ${\bf %
r}_{1}$ an associated bulk system with a uniform density 
\begin{equation}
\tilde{\rho}^{{\bf r}_{1}}\equiv \rho (\tilde{\mu}^{{\bf r}_{1}}),
\label{eq:rhorr}
\end{equation}
whose correlation functions are used in the expansion. Equation \ref
{eq:sumrunexpand} thus becomes 
\begin{eqnarray}
\rho ({\bf r}_{1}) &=&\Lambda ^{-3}e^{\beta [\tilde{\mu}^{{\bf r}_{1}}-%
\tilde{\phi}^{{\bf r}_{1}}({\bf r}_{1})]+c^{(1)}(\tilde{\rho}^{{\bf r}_{1}})}
\nonumber \\
&&\times \{ 1+\int d{\bf r}_{2}\,c^{(2)}(r_{12};\tilde{\rho}^{{\bf r}%
_{1}})(\rho ({\bf r}_{2})-\tilde{\rho}^{{\bf r}_{1}})   \nonumber \\
&& +Q({\bf r}_{1})+...\} 
\label{eq:sumruleexpandr}
\end{eqnarray}

In principle (assuming convergence of the infinite series), if we could
accurately evaluate {\em all} terms in the Taylor series, the same exact
result for $\rho ({\bf r}_{1})$ would be found for any choice of $a^{{\bf r}%
_{1}}$ due to the shifting property. But this is hopelessly complicated in
general and approximate values for $\rho ({\bf r}_{1})$ from truncations of
the series do depend on the choice of the $a^{{\bf r}_{1}}.$

\section{Shifted linear truncations}

\label{sec:SLR}Our strategy is to try to choose the $a^{{\bf r}_{1}}$ or the 
$\tilde{\rho}^{{\bf r}_{1}}$ at {\em each} ${\bf r}_{1}$ in such a way that
a self consistent solution for $\rho ({\bf r}_{1})$ based on a simple low
order truncation of the series in eq \ref{eq:sumrunexpand} can give accurate
results. In particular, we suppose that the $a^{{\bf r}_{1}}$ can be chosen
by some argument to be specified later so that the expansion can be
truncated at linear order to a good approximation. We thus arrive at a very
general starting point, which we refer to as the {\em shifted linear
response }(SLR) equation: 
\begin{eqnarray}
\rho ({\bf r}_{1}) &=&\Lambda ^{-3}e^{\beta [\tilde{\mu}^{{\bf r}_{1}}-%
\tilde{\phi}^{{\bf r}_{1}}({\bf r}_{1})]+c^{(1)}(\tilde{\rho}^{{\bf r}_{1}})}
\nonumber \\
&&\times \left[ 1+\int d{\bf r}_{2}\,c^{(2)}(r_{12};\tilde{\rho}^{{\bf r}%
_{1}})(\rho ({\bf r}_{2})-\tilde{\rho}^{{\bf r}_{1}})\right] ~~~ \nonumber \\
&=&\tilde{\rho}^{{\bf r}_{1}}e^{-\beta \tilde{\phi}^{{\bf r}_{1}}({\bf r}%
_{1})}  \nonumber \\
&&\times \left[ 1+\int d{\bf r}_{2}\,c^{(2)}(r_{12};\tilde{\rho}^{{\bf r}%
_{1}})(\rho ({\bf r}_{2})-\tilde{\rho}^{{\bf r}_{1}})\right] ~~~
\label{eq:masterequation}
\end{eqnarray}

Specific choices of the $a^{{\bf r}_{1}}$ or $\tilde{\rho}^{{\bf r}_{1}}$
will lead to different approximations for $\rho ({\bf r}_{1}).$ The PY
approximation arises from the choice $a^{{\bf r}_{1}}=0$ or $\tilde{\rho}^{%
{\bf r}_{1}}=\rho ^{B}.$ As we argued above this choice should give very
accurate results for sufficiently localized fields and it is known to give a
good description of the response to hard core solutes.

However, when the external field is extended but slowly varying, a different
choice of $a^{{\bf r}_{1}}$ is clearly more appropriate. As discussed in
detail in References \onlinecite{weeks02}, \onlinecite{katsov01}, and
\onlinecite{chen04}, the HLR equation is very accurate
in such cases. This uses the {\em hydrostatic shift, }where the
external field is locally shifted at each ${\bf r}_{1}$ to be zero. This
corresponds to the choice 
\begin{equation}
a^{{\bf r}_{1}}=\phi ({\bf r}_{1}),  \label{eq:ahydrostaticshift}
\end{equation}
in eq \ref{eq:masterequation} so that $\tilde{\phi}^{{\bf r}_{1}}({\bf r}
_{1})=0$. The associated uniform density given by the shifted chemical
potential for this particular choice of $a^{{\bf r}_{1}}$ is denoted $\rho ^{%
{\bf r}_{1}}$ and it satisfies 
\begin{equation}
\rho ^{{\bf r}_{1}}=\rho (\mu ^{B}-\phi ({\bf r}_{1})).
\label{eq:hydrostaticrho}
\end{equation}

Unlike the PY approximation, the HLR equation builds in the invariance of
the grand canonical ensemble when both the chemical potential and the
external field are shifted by a constant. It is an excellent approximation
when the external field is slowly varying, and the expansion in eq \ref
{eq:sumruleexpandr} about the hydrostatic density eq \ref{eq:hydrostaticrho}
converges rapidly.\cite{localdensity}
Moreover the HLR equation gives the same results as the
PY equation for hard core fields. In general the HLR approximation is much
more accurate than the PY approximation for a wide range of external fields.%
\cite{katsov01,chen04}

The SLR equation \ref{eq:masterequation} with the particular choice $a^{{\bf %
r}_{1}}=\phi ({\bf r}_{1})$ thus provides an alternative derivation of the
HLR equation, in addition to the previous physical arguments based on
optimizing linear response by use of the hydrostatic shift.\cite
{weeks02,katsov01,chen04} However, as discussed below, there are some limits
where the HLR choice for $a^{{\bf r}_{1}}$ gives poor results. The SLR
equation permits other choices for $a^{{\bf r}_{1}}$ and thus provides
additional flexibility that can lead to improvements of the HLR equation.
With proper choice of $a^{{\bf r}}$, eq \ref{eq:masterequation} can be used
to describe and bridge several limits that both the PY and the HLR equations
fail to capture. These limiting cases will be addressed in the following.

\subsection{{}Tiny Fields}

Consider first a very localized field $\phi ({\bf r})$ that is non-vanishing
only within a spherical volume $V^{\phi }$ of radius $d/2$, with $d$ the
hard sphere diameter:

\begin{eqnarray}
\phi ({\bf r}) &\neq &0,\,|{\bf r}|<\frac{d}{2}  \nonumber \\
&=&0,\,{\rm otherwise}.  \label{eq:tiny field}
\end{eqnarray}
This volume is so small that it can simultaneously accommodate the centers
of at most {\em one} hard sphere particle. We call localized fields that are
nonzero only in such a tiny region {\em tiny fields}. A special case is a 
{\em tiny cavity}, where $\phi $ is infinite in $V^{\phi }.$ The density
response of a hard sphere fluid to any tiny field can be determined exactly,
as we now show.

We start with the grand partition function $\Xi [\phi ]$ when a general
external field $\phi $ is present in the fluid: 
\begin{eqnarray}
\Xi [\phi ] &=&\sum_{N=0}^{\infty }\frac{z^{N}}{N!}Z_{N}[\phi ]  \nonumber \\
&=&\sum_{N=0}^{\infty }\frac{z^{N}}{N!}\int d{\bf r}_{1}...{\bf r}%
_{N}e^{-\beta \sum\limits_{n=1}^{N}\phi ({\bf r}_{i})-\beta V_{N}(\{{\bf r}%
\})}.  \label{eq:Xiphi}
\end{eqnarray}
Here $Z_{N}[\phi ]$ is the canonical partition in the presence of the field, 
$Z_{N}[0]$ refers to that of the uniform fluid when the external field is
zero, $z\equiv \exp (\beta \mu ^{B})/\Lambda ^{3}$, and $V_{N}(\{{\bf r}\})$
is the intermolecular interaction potential
between the $N$ fluid particles. Introducing the Mayer $f$-function for the
external field 
\begin{equation}
f^{\phi }({\bf r})\equiv e^{-\beta \phi ({\bf r})}-1,  \label{eq:mayerf}
\end{equation}
eq \ref{eq:Xiphi} can be rewritten in terms of an expansion about the
uniform fluid with $\phi =0$:
\begin{widetext}
\begin{eqnarray}
\Xi [\phi ] &=&\sum_{N=0}^{\infty }\frac{z^{N}}{N!}\sum_{n=0}^{N}\int d{\bf r%
}_{1}d{\bf r}_{2}...d{\bf r}_{n}\frac{N!}{(N-n)!n!}f^{\phi }({\bf r}%
_{1})f^{\phi }({\bf r}_{2})...f^{\phi }({\bf r}_{n})\int d{\bf r}_{n+1},...,d%
{\bf r}_{N}e^{-\beta V_{N}(\{{\bf r}\})}  \nonumber \\
&=&\sum_{n=0}^{\infty }\frac{1}{n!}\int d{\bf r}_{1}d{\bf r}_{2}...d{\bf r}%
_{n}f^{\phi }({\bf r}_{1})f^{\phi }({\bf r}_{2})...f^{\phi }({\bf r}%
_{n})\sum_{N\geq n}^{\infty }\frac{z^{N}}{(N-n)!}\int d{\bf r}_{n+1},...,d%
{\bf r}_{N}e^{-\beta V_{N}(\{{\bf r}\})}  \nonumber \\
&=&\Xi [0]\sum_{n=0}^{\infty }\frac{1}{n!}\int d{\bf r}_{1}d{\bf r}_{2}...d%
{\bf r}_{n}f^{\phi }({\bf r}_{1})f^{\phi }({\bf r}_{2})...f^{\phi }({\bf r}
_{n})\rho ^{(n)}({\bf r}_{1},...,{\bf r}_{n};[0])  \label{eq:Xiphifexpand}
\end{eqnarray}
\end{widetext}
Here $\rho ^{(n)}({\bf r}_{1},...,{\bf r}_{n};[0])$ is the $n$-particle
distribution function in the uniform grand canonical ensemble with chemical
potential $\mu ^{B}.$

For general extended fields this formal expansion does not converge rapidly.
However it can be very useful when the field is confined to a small
localized region of space since $f^{\phi }({\bf r)}$ is nonzero only where
$\phi $ is nonzero. In particular for tiny fields the expansion eq \ref
{eq:Xiphifexpand} must truncate {\em exactly} due to the vanishing of the
$\rho ^{(n)}({\bf r}_{1},...,{\bf r}_{n};[0])$ when more than one hard
particle is simultaneously within the volume $V^{\phi }$. By functional
differentiation of eq \ref{eq:Xiphifexpand} we can also find an expansion
for $\rho ({\bf r};[\phi ])$ that similarly truncates. Thus we have exactly
for tiny fields 
\begin{equation}
\Xi [\phi ]=\Xi [0]\left[ 1+\int d{\bf r}_{1}\rho ^{B}f^{\phi }({\bf r}%
_{1})\right] ,  \label{eq:Xiphitinty}
\end{equation}
and 
\begin{eqnarray}
\rho ({\bf r}_{1};[\phi ]) &=&\frac{\rho ^{B}e^{-\beta \phi ({\bf r}_{1})}}{%
1+\int d{\bf r}_{2}\rho ^{B}f^{\phi }({\bf r}_{2})}  ~~~~~ \nonumber \\
&&\,\,\,\times \left[ 1+\int d{\bf r}_{2}\rho ^{B}g^{(2)}(r_{12};
\rho ^{B})f^{\phi }({\bf r}_{2})\right] .~~~~~  \label{eq:rhophitinty}
\end{eqnarray}
The exact result for the density response to a tiny cavity was originally
derived in a different way by Reiss and Casberg.\cite{Reisstinyfield} Here
$g^{(2)}(r_{12};\rho ^{B})$ is the exact radial
distribution for the uniform hard sphere fluid. This same formula clearly
holds for a more general model with longer ranged pair interactions outside
the hard core if the appropriate $g^{(2)}$ is used.

As would be expected by the appearance of $\rho ^{B}$ in this equation, one
can show that the PY approximation eq \ref{eq:PY} is consistent with this
exact result for {\em any} tiny field. See reference \onlinecite{chenthesis} for a
straightforward but tedious derivation. In accord with the qualitative
argument above, the linear extrapolation from the bulk into the tiny region
is {\em exact} in this case. This corresponds in the SLR equation to the
choice $a^{{\bf r}}=0$, i.e., $\tilde{\rho}^{{\bf r}}=\rho ^{B}$ and shows
that the series indeed truncates exactly in this special case.

However, a closely related limit highlights a general problem with the PY
approximation. Consider a field $\phi ^{c}({\bf r})$ that is a non-zero
constant $c$ outside a tiny region $V^{\phi }$ and any value $\phi ({\bf r})$
inside. This can immediately be shifted to be the type for which eqs \ref
{eq:Xiphitinty} and \ref{eq:rhophitinty} hold by making the choice $a^{r}=c.$
For such a field $\phi ^{c}$ we then have 
\begin{eqnarray}
\rho ({\bf r}_{1};[\phi ]) &=&\frac{\tilde{\rho}^{c}e^{-\beta \tilde{\phi}^{c}({\bf r})}}{%
1+\int d{\bf r}_{2}\tilde{\rho}^{c}\tilde{f}^{c}({\bf r}_{2})} ~~~~ \nonumber \\
&&\,\,\,\times \left[ 1+\int d{\bf r}_{2}\tilde{\rho}^{c}g^{(2)}(r_{12};
\tilde{\rho}^{c})\tilde{f}^{c}({\bf r}_{2})\right] .~~~~  \label{eq:rhotinyshift}
\end{eqnarray}
Here $\tilde{\rho}^{c}=\rho (\mu ^{B}-c)$ and $\tilde{f}^{c}$ is defined as
in eq \ref{eq:mayerf} with the shifted tiny field $\tilde{\phi}^{c}({\bf r}%
)=\phi ^{c}({\bf r})-c$.

For a perturbation that varies significantly only in a very local region
(compared with the correlation length of the particles), a particle situated
in the perturbed region, though directly affected by the field, should
screen the perturbation from the rest of the fluid. The fluid's response
thus essentially remains that of the uniform fluid outside the local region.
The shifted field represents such a localized perturbation. Thus choosing
the density $\tilde{\rho}^{c}$ to be that of the bulk environment, i.e., $%
\rho (\mu ^{B}-c),$ will truncate eq \ref{eq:Xiphifexpand} at low order,
leading to eq \ref{eq:rhotinyshift}. And again, the choice $a^{{\bf r}}=c$
in the SLR equation \ref{eq:masterequation} gives this exact result.

In this essentially equivalent case however, the PY approximation, which
always uses the {\em unshifted} $\rho ^{B},$ will give an incorrect result,
even though it can exactly describe the tiny field limit when $c=0.$ Unlike
the SLR equation, the PY approximation does not build in the exact shifting
feature of the grand canonical ensemble. This can cause significant errors
for extended slowly varying fields.

However, the HLR approximation uses the hydrostatic shift eq \ref
{eq:ahydrostaticshift} and thus will correctly describe the shifted bulk
density $\tilde{\rho}^{c}$ in this case. Moreover it is exact for {\em tiny
cavities} (tiny fields that are infinitely repulsive inside $V^{\phi }$)
since any finite value for $a^{{\bf r}}$ inside the cavity will still give
the correct zero density. However the HLR equation is {\em not} exact for
more general finite tiny fields. The HLR reference density $\rho ^{{\bf r}}$
would follow the variations in $\phi $ inside the tiny region, contrary to
the exact result with a constant $\tilde{\rho}^{c}$ everywhere. Rapidly
varying tiny fields can thus generate noticeable errors in the HLR
approximation, as will be shown in a later section where computational
results are reported.

\subsection{Narrow Slits}

Another application that may at first seem to be very different from the
tiny field case is when an extended external field confines the fluid to a
region of reduced dimensions. For example, consider a hard sphere fluid
confined between two planar hard walls forming a narrow slit. We can get
exact results for this case from eq \ref{eq:Xiphifexpand} by exploiting the
shifting property of the grand ensemble. The confining field can be taken to
be the limit of piecewise constant potentials defined so that 
\begin{eqnarray}
\phi ^{c}(z) &=&0,\,\,\,0<z<L_{s}  \nonumber \\
&=&c,\,\,\,\text{otherwise},  \label{eq:2wall}
\end{eqnarray}
in the limit where $c\rightarrow \infty .$ In this limit the fluid's density
will be zero except in the narrow region between the walls. Here $L_{s}$ is
the effective width of the slit as seen by the centers of the fluid
particles.

If we formally introduce the {\em uniform} shift $a^{z}=c$ we have 
\begin{eqnarray}
\tilde{\phi}^{c}(z) &=&-c,\,\,0<\,z<L_{s}  \nonumber \\
&=&0,\,\text{otherwise}  \label{eq:shiftphislit}
\end{eqnarray}
and the shifted external field $\tilde{\phi}^{c}(z)$ is non-zero only in the
narrow slit region, similar to a tiny field. However, the shifted slit field
is not strictly a tiny field as defined in the previous section where the
expansion exactly truncates, because even when $L_{s}\rightarrow 0$, many
particles in principle can still be found in the slit, aligned in a
two-dimensional layer along the walls of the slit. But once this shift has
been made, the expansion in eqs \ref{eq:Xiphifexpand} and \ref
{eq:rhotinyshift} converges rapidly for small $L_{s}$ since the
contributions from the integration over the $\tilde{f}^{c}$ tend to zero.
The shifted chemical potential $\tilde{\mu}^{c}=\mu ^{B}-c$ tends to $%
-\infty ,$ corresponding to an expansion about the ideal gas limit where the
shifted bulk density $\tilde{\rho}^{c}$ tends to zero and $c^{(1)}(\tilde{%
\rho}^{c})=0.$

Inside the slit where $\tilde{\mu}^{c}-\tilde{\phi}^{c}({\bf r}_{1})=\mu
^{B}-\phi ^{c}({\bf r}_{1})=\mu ^{B}$, we have a finite limiting density as $%
L_{s}\rightarrow 0$ given by 
\begin{equation}
\tilde{\rho}^{c}e^{-\beta \tilde{\phi}^{c}({\bf r}_{1})}=\Lambda
^{-3}e^{\beta [\tilde{\mu}^{c}-\tilde{\phi}^{c}({\bf r}_{1})]+c^{(1)}(\tilde{%
\rho}^{c})}=\Lambda ^{-3}e^{\beta \mu ^{B}}.  \label{eq:limitingrhoinslit}
\end{equation}
Equation \ref{eq:rhotinyshift} then gives the first two terms in an exact
(but non-truncating) virial-like expansion valid for narrow slits. Higher
order terms can be determined straightforwardly from eq \ref{eq:Xiphifexpand}%
. In agreement with previous work, \cite{hendersonslit} there is a constant
limiting lowest order density profile in the narrow slit given by 
\begin{eqnarray}
\rho (z) &=&\Lambda ^{-3}e^{\beta \mu ^{B}},\,\,0<z<L_{s}  \nonumber \\
&=&0,\,\,\,{\rm otherwise.}  \label{eq:exactrhoslithenderson}
\end{eqnarray}
For a value of $\mu ^{B}$ corresponding to a dense uniform hard sphere
fluid, this yields a very large limiting value for the reduced density in
the slit $\rho ^{3D}d^{3}\equiv \Lambda ^{-3}e^{\beta \mu ^{B}}d^{3}\gg 1$.
However the density of fluid particles per unit area of the wall $\rho
^{2D}d^{2}\equiv \Lambda ^{-3}e^{\beta \mu ^{B}}d^{2}L_{s}$ tends to zero as 
$L_{s}\rightarrow 0.$ Thus particles in the narrow slit are very far apart
laterally and an expansion about the ideal gas limit is physically
appropriate.

Clearly the SLR equation can reproduce these exact limiting results if the
proper choice $a^{z}=c$ (where $c\rightarrow \infty $) for all $z$ is made.
However, the PY approximation uses $a^{z}=0$ everywhere in the SLR equation,
while the HLR equation assumes $a^{z}=c$ outside the slit but $a^{z}=0$
inside, and hence they both give incorrect results in the limit $%
L_{s}\rightarrow 0$. Both theories correctly predict zero density outside
the slit, since for any choice of $a^{z}$, the factor $\tilde{\rho}%
^{z_{1}}e^{-\beta \tilde{\phi}^{z_{1}}(z_{1})}$ in eq \ref{eq:masterequation}
immediately makes $\rho (z_{1})$ zero outside the slit. However, inside the
slit both the PY and the HLR equations take $\tilde{\rho}^{z_{1}}=\rho ^{B}$
and thus expand about the uniform bulk reference state, which gives a very
poor description of the dilute 2D gas in the slit. As $L_{s}\rightarrow 0$
they predict a much lower limiting density \cite{slitexample} than given by
eq \ref{eq:exactrhoslithenderson}. These problems arise only at small
separations of the order or smaller that of the hard sphere diameter $d$. At
larger separations and in the one wall limit both theories give much more
satisfactory results.

\section{A new criterion for choosing the reference density in the SLR
equation}

\label{sec:new criterion}The above discussion has shown the versatility of
the SLR equation and its ability to give exact limiting results in several
specific cases with proper choice of the $\tilde{\rho}^{{\bf r}}.$ It has
also shown that inaccuracies arise in some cases from the prescribed local
choices made by the PY and HLR equations. Thus we need a more general and
systematic way to choose $\tilde{\rho}^{{\bf r}}$ in the SLR equation. To
that end we first look more closely at the reasons why HLR fails in some
cases.

\subsection{Limitations of the HLR prescription}

The HLR equation expands about the hydrostatic density $\rho ^{{\bf r}}=\rho
(\mu ^{B}-\phi ({\bf r})).$ This depends on the external field too locally
in cases where the external fields varies significantly in local regions
much smaller than the correlation length of the fluid. In such cases, the
proper density $\tilde{\rho}^{{\bf r}}$ to expand about is often not the
hydrostatic density, but a nonlocal extrapolation using the density of the
surroundings, as illustrated by the tiny field and narrow slit examples
discussed above. To use the SLR equation \ref{eq:masterequation} to improve
on the HLR approximation, we need a new way to choose $\tilde{\rho}^{{\bf r}%
} $ that can account for this extrapolation of the local uniform system in
such cases, while not spoiling the good results of the HLR equation in most
other limits. We describe below our first attempt to develop such a
computationally useful criterion.

\subsection{The {}insensitivity criterion}

If all terms are exactly retained in eq \ref{eq:sumruleexpandr}, it should
be invariant with respect to a simultaneous shift of the chemical potential
$\mu ^{B}$ and the external field $\phi $. Thus eq \ref{eq:sumruleexpandr}
should hold for all choices of $\tilde{\rho}^{{\bf r}}$. However only
certain choices of $\tilde{\rho}^{{\bf r}}$ can efficiently truncate the
series at low orders. One possible criterion for a truncation is to choose
$\tilde{\rho}^{{\bf r}}$ that minimize the contribution from the quadratic
term $Q({\bf r}_{1})$
in eq \ref{eq:sumruleexpandr}. However, unlike $c^{(2)}(r_{12};\rho )$,
$c^{(3)}({\bf r}_{1},{\bf r}_{2},{\bf r}_{3};\rho )$ is often not available
analytically (and accurately) and the 6-dimensional integral of
$Q({\bf r}_{1})$ in eq. \ref{eq:Qexpand} is very computational demanding.

To circumvent the difficulty of dealing with $Q({\bf r})$ directly, a
reasonable alternative is to consider how the predictions of the SLR
equation change as $\tilde{\rho}^{{\bf r}}$ is varied. Since the SLR
equation is a truncation of the exact series in eq \ref{eq:sumruleexpandr},
it is certainly not invariant with respect to variation of {\em any} $\tilde{%
\rho}^{{\bf r}}$. However if the truncation is accurate for some particular
choice of $\tilde{\rho}^{{\bf r}},$ then in effect the higher order terms in
the series have been taken into account. Thus the SLR equation should be
relatively {\em insensitive} to small variations about the particular $%
\tilde{\rho}^{{\bf r}}$ that make the higher order corrections to the SLR
equation small. This condition need not be exact, even in the case of a tiny
field where the series truncates exactly, but it seems likely that it could
produce reasonable choices for $\tilde{\rho}^{{\bf r}}$ in many cases.

This leads to the following self-consistent condition for the density given
by the SLR equation \ref{eq:masterequation}: 
\begin{equation}
\delta \rho ({\bf r}_{1})/\delta \tilde{\rho}^{{\bf r}_{2}}=0,\,\forall {\bf %
r}_{1},{\bf r}_{2},  \label{eq:mhlrjdwdiff}
\end{equation}
expressing the insensitivity of the density with respect to variations in
$\tilde{\rho}^{{\bf r}}$. Differentiating both sides of the SLR equation \ref
{eq:masterequation} and collecting the terms arising from $\delta \rho ({\bf %
r}_{1})/\delta \tilde{\rho}^{{\bf r}_{2}}$ (for details of the derivation,
see the appendix), the {\em insensitivity criterion} (IC) arising from eq 
\ref{eq:mhlrjdwdiff} can be written as: 
\begin{equation}
\tilde{\rho}^{{\bf r}_{1}}=\frac{\int d{\bf r}_{2}W(|{\bf r}_{1}-{\bf r}%
_{2}|;\tilde{\rho}^{{\bf r}_{1}})\rho ({\bf r}_{2})}{\int d{\bf r}_{2}W(|%
{\bf r}_{1}-{\bf r}_{2}|;\tilde{\rho}^{{\bf r}_{1}})}.
\label{eq:mhlrjdwcriteriaave}
\end{equation}
where 
\begin{equation}
W(r_{12};\rho )\equiv \dot{c}^{(1)}(\rho )c^{(2)}(r_{12};\rho )+
\dot{c}^{(2)}(r_{12};\rho )
\label{eq:theWjdw}
\end{equation}
and 
\begin{equation}
\dot{c}^{(1)}(\rho )\equiv dc^{(1)}(\rho )/d\rho
;\,\,\,\,\dot{c}^{(2)}(r_{12};\rho )
\equiv dc^{(2)}(r_{12};\rho )/d\rho .  \label{eq:cdot}
\end{equation}
Because the function $W(|{\bf r}_{1}-{\bf r}_{2}|;\rho )$ in eq \ref
{eq:theWjdw} has range of $c^{(2)}(|{\bf r}_{1}-{\bf r}_{2}|;\rho )$, eq \ref
{eq:mhlrjdwcriteriaave} shows that $\tilde{\rho}^{{\bf r}_{1}}$ can be
interpreted as the full density $\rho ({\bf r})$ averaged over the range of
the fluid's correlation length around the point ${\bf r}_{1}$, using a
self-consistent {\em weighting function} $W$ that itself depends on $\tilde{%
\rho}^{{\bf r}_{1}}$. We will refer to the resulting $\tilde{\rho}^{{\bf r}}$
as the {\em smoothed reference density} in what follows. Some versions of
weighted density functional theory have used similar weighted densities,
though the detailed implementation and justification are rather different.%
\cite{evans92,meister85}

Equation \ref{eq:mhlrjdwcriteriaave} derived from the IC\ can then be solved
along with the SLR equation to determine both the full density $\rho ({\bf r})$
and the smoothed density $\tilde{\rho}^{{\bf r}}$. We refer to these coupled
equations as the {\em IC equations}. The IC equations can be solved
numerically by iteration with the same methods used to solve the PY or HLR
equations.

\subsection{Behavior of IC equations in limiting cases}

We first verify that the IC equations can give accurate results in limiting
cases where the proper choice of $\tilde{\rho}^{{\bf r}}$ is known. In the
hydrostatic limit where the external field is very slowly varying, $\rho (%
{\bf r})\ $will reduce to the hydrostatic density $\rho ^{{\bf r}}=\rho (\mu
^{B}-\phi ({\bf r}))$, as given by the HLR equation. In this same limit $%
\rho ({\bf r}_{2})$ in the IC equation \ref{eq:mhlrjdwcriteriaave} can be
approximated by $\rho ({\bf r}_{1})$ and taken outside the integral. This
gives $\tilde{\rho}^{{\bf r}}=\rho ({\bf r})$ and hence $\tilde{\rho}^{{\bf r%
}}=\rho ^{{\bf r}}.$ The IC equations thus reduce to the HLR equation for
slowly varying fields and recover the hydrostatic limit correctly. However,
because of the averaging in eq \ref{eq:mhlrjdwcriteriaave}, in other limits
the IC choice of $\tilde{\rho}^{{\bf r}}$ is less local than the HLR choice $%
\rho ^{{\bf r}}$ and tends to smear out the nonuniformity caused by external
perturbations in small regions.

\begin{figure}[tbp]
\includegraphics[%
  width=0.93\columnwidth]{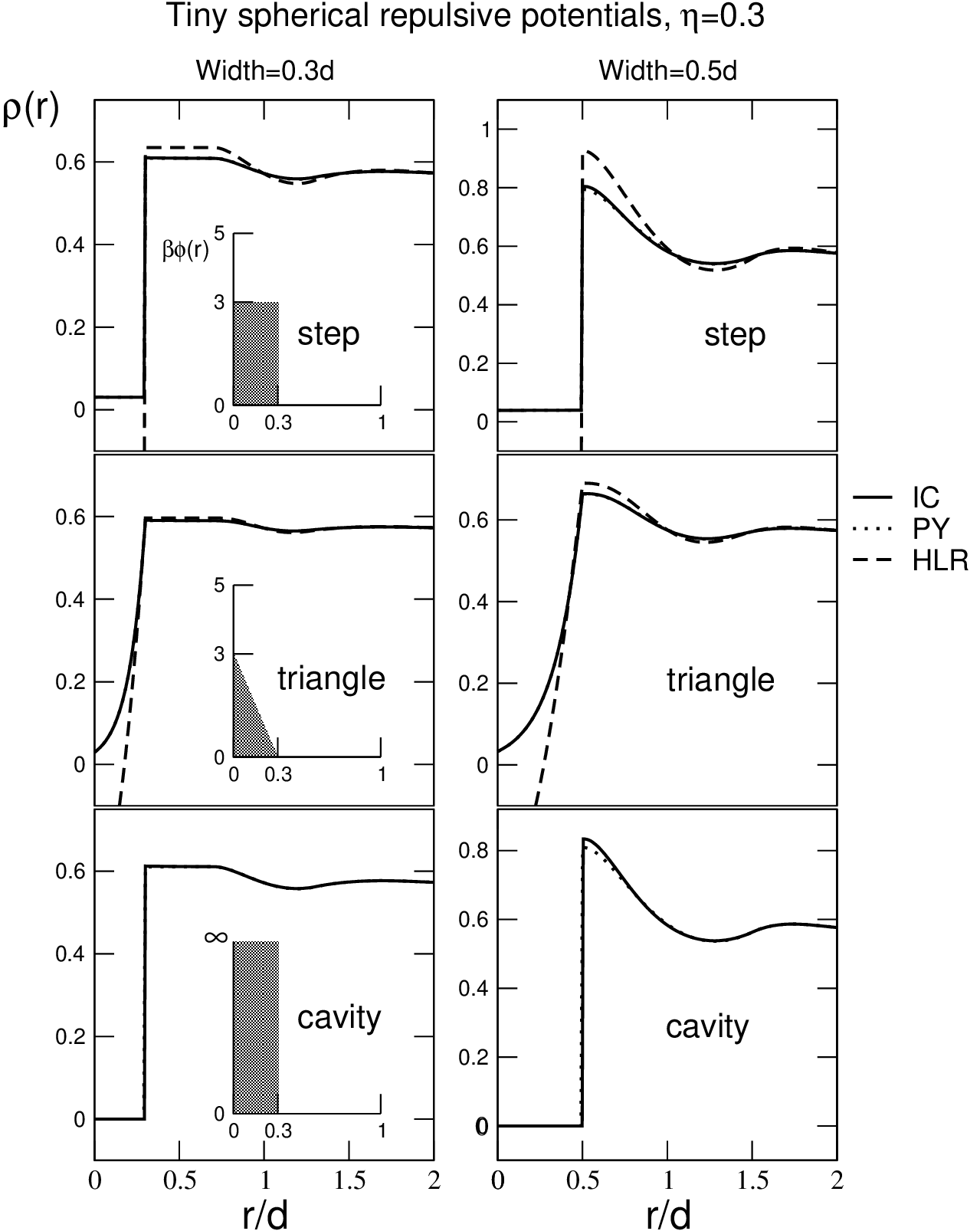}

\caption{The density response to tiny repulsive external fields of different
widths $W$ is plotted. The curves are the predictions by the IC, PY and HLR
approximations. The PY approximation is exact for the density values inside
the non-zero field region. All the external fields are spherical. {}``Step''
denotes a spherical step function where $\beta \phi (r)=3,\forall r<$ $W$
and $\beta \phi (r)=0,\text{ otherwise}$. {}``Triangle'' refers to the
potential $\beta \phi (r)=3-3r/W$ that has the same height as the step
potentials but decays linearly to zero at $r=W$ with $\beta \phi (r)=0,\text{
otherwise}$. {}``Cavity'' refers to the hard core potential $\beta \phi
(r)=\infty ,\forall r<W$ and $\beta \phi (r)=0,\text{ otherwise}$. The form
of the potentials $\beta \phi (r)$ are illustrated in the insets. For the
cavity potentials, the PY and HLR approximations give identical density
solutions. The bulk fluid's packing fraction $\eta \equiv \pi \rho
^{B}d^{3}/6$ is $\eta =0.3$.}
\label{cap:tinyreplusive}
\end{figure}

For the narrow slit limit discussed above, the $\tilde{\rho}^{{\bf r}}$
given by eq \ref{eq:mhlrjdwcriteriaave} correctly approaches zero as $%
L\rightarrow 0$, since the Boltzmann factor $e^{-\beta \phi ({\bf r})}$
ensures that $\rho ({\bf r})$ is zero inside the walls. Thus the IC
equations are {\em exact} in the narrow slit limit as $L_{s}\rightarrow 0$
and correct the poor predictions of both the HLR and PY equations.

\begin{figure}[tbp]
\includegraphics[%
  width=0.93\columnwidth]{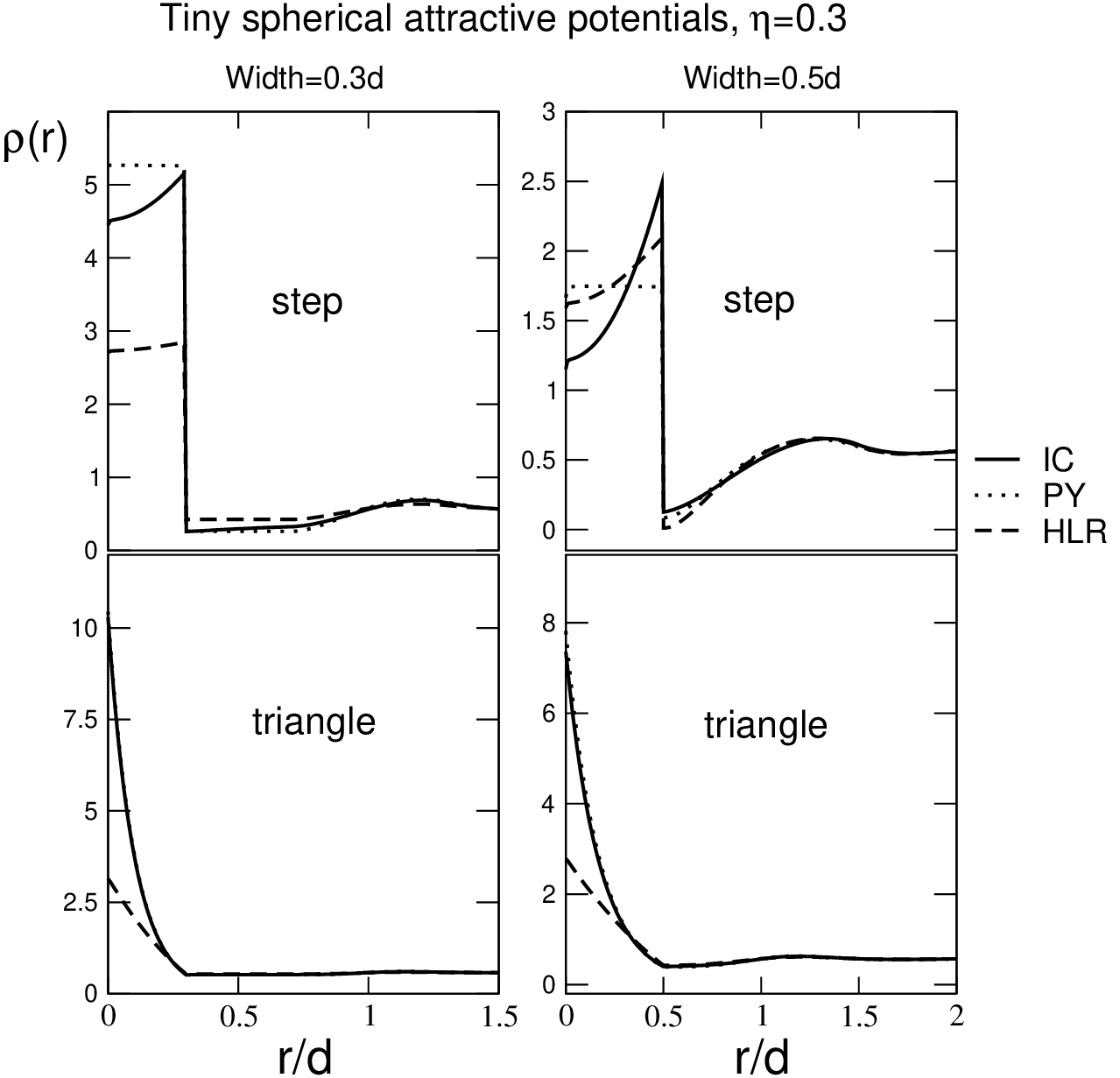}
  
\caption{The curves plotted here follow the same legend conventions used in
Figure \ref{cap:tinyreplusive} and are computed at the same bulk packing
fraction $\eta =0.3$. The external fields calculated here are all tiny
spherical attractive potentials. Step and triangle potentials are defined as
in Figure \ref{cap:tinyreplusive}, except that the sign of the potential is
negative.}
\label{cap:tinyattractive}
\end{figure}

For tiny fields, the IC choice in eq \ref{eq:mhlrjdwcriteriaave} strictly
reproduces the exact bulk density choice $\tilde{\rho}^{{\bf r}}=\rho (\mu
^{B}-\phi (\infty ))$ only in the limit where $V^{\phi }\rightarrow 0,$ and
is not exact for general {}tiny fields. However $\tilde{\rho}^{{\bf r}}$ is
generally very close to that of the bulk density because the tiny field
volume $V^{\phi }$ has little weight in the averaging. Thus the IC equations
can be expected to agree reasonably well with exact results for general tiny
fields, as will be shown in the next section.

\section{Numerical results}

\label{sec:results} 

We first consider the density response of a hard sphere fluid at a moderate
packing fraction $\eta =0.3$ to a series of spherical model potentials of
varying ranges and different signs. In particular we studied repulsive
(attractive) step functions of height $3k_{B}T$ ($-3k_{B}T$) and different
ranges and ``triangular'' fields that start with the same height at the
origin and vary linearly in $r$ to the cutoff. Hard sphere cavities with the
same cutoffs were also studied. Numerical solutions of the IC, HLR, and PY
equations are presented, together with results of Monte Carlo simulations
for the fluid's density response. The PY bulk direct correlation function $%
c^{(2)}(r_{12};\rho )$, which is very accurate at these densities, was used
in the theoretical calculations.\cite{hansenmac} The Carnahan-Starling \cite
{CSeos} equation of state was used for computing the density $\tilde{\rho}^{%
{\bf r}}=\rho (\mu ^{B}-a^{{\bf r}})$ of the locally shifted uniform system.

\subsection{Tiny fields}

\begin{figure}[tbp]
\includegraphics[%
  width=0.93\columnwidth]{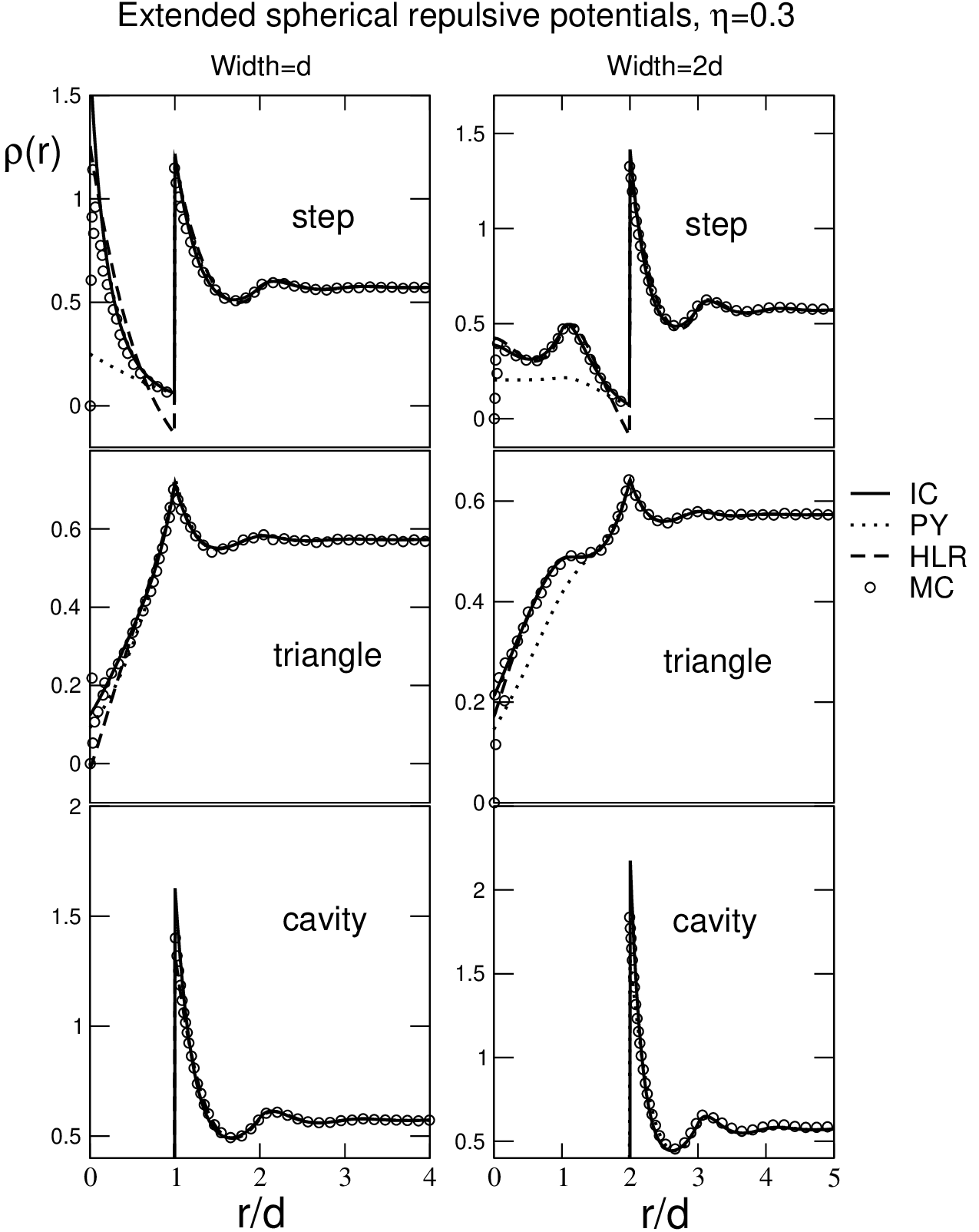}
  
\caption{Density response to extended spherical repulsive potentials of
varying widths. Conventions and bulk packing fraction are those of Figures 
\ref{cap:tinyreplusive} and \ref{cap:tinyattractive}.}
\label{cap:nontinyreplusive}
\end{figure}

For tiny fields, all results should be compared to the PY approximation,
which is exact for such fields (subject only to the very small errors in the
PY bulk direct correlation function). As can be seen in Figures \ref
{cap:tinyreplusive} and \ref{cap:tinyattractive}, the HLR equation is exact
only for tiny {\em cavities}. For finite tiny fields, its major errors occur
in the tiny region where the field is nonvanishing and rapidly varying. The
density response predicted by the HLR equation often exhibits a negative
region where the external field varies most rapidly. The IC approximation,
on the other hand, in general agrees with the PY approximation much better
in the tiny field region and in particular eliminates the negative densities
given by HLR. However the IC equations are not exact for tiny fields, and
tend to overestimate the contact densities.

\subsection{Extended fields}

\begin{figure}[tbp]
\includegraphics[%
  width=0.93\columnwidth]{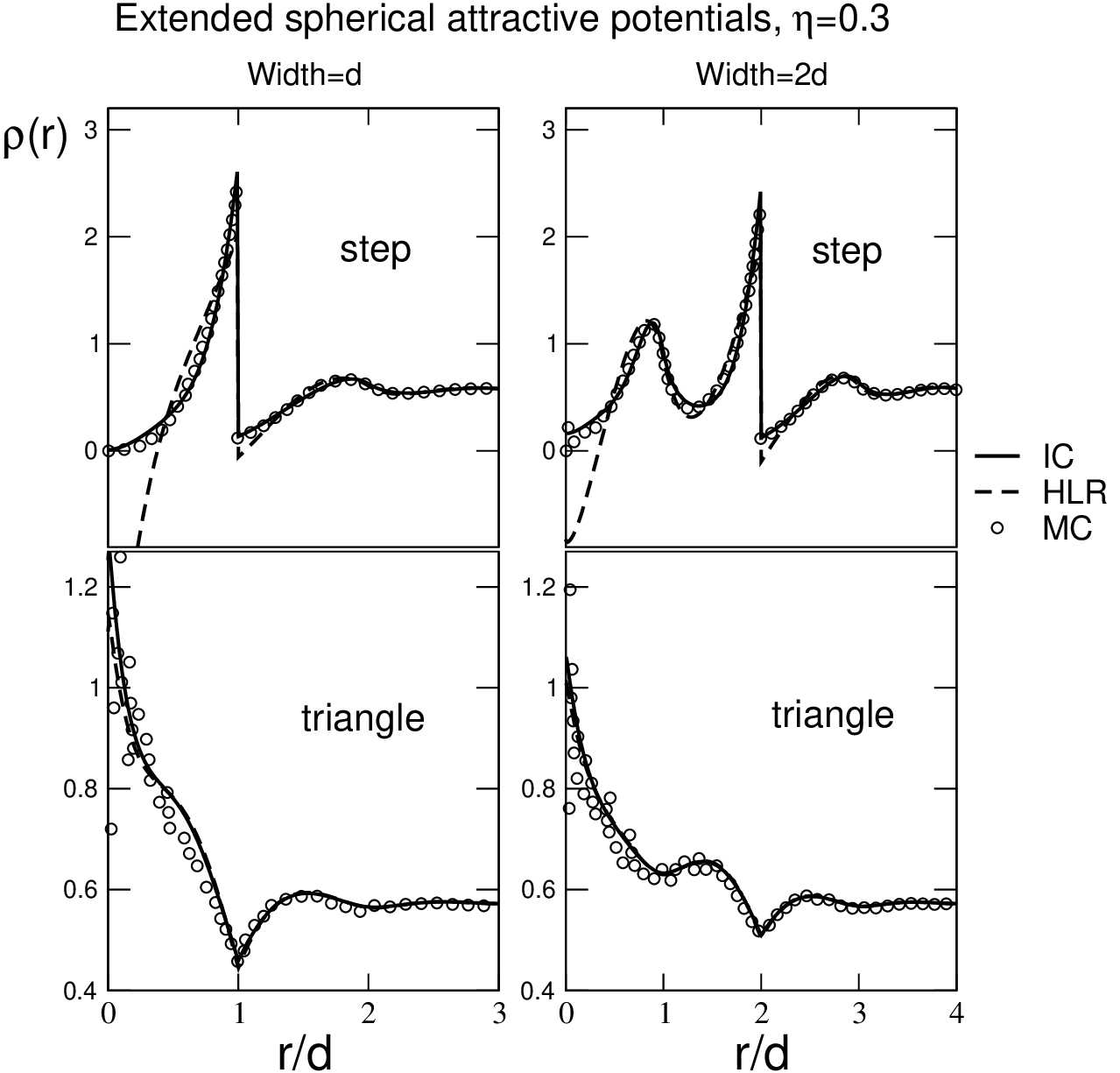}

\caption{Density response to extended spherical attractive potentials of
varying widths. Conventions and bulk packing fraction are those of Figures 
\ref{cap:tinyreplusive} and \ref{cap:tinyattractive}.}
\label{cap:nontinyattractive}
\end{figure}

For more extended fields, none of the approximations are exact, so Monte
Carlo simulations were carried out to test the various theories. As shown in
Figures \ref{cap:nontinyreplusive} and \ref{cap:nontinyattractive}, as the
range of the step and the triangle potentials becomes wider, the HLR
approximation becomes more accurate. However, it can still exhibit negative
densities in its solution for both repulsive and attractive step potentials,
especially for the narrower steps. For the same width of the potentials, the
HLR equation does better in predicting the response to the triangle
potentials than to the step potentials. This agrees with the expectation
that the HLR equation should be more accurate when the external field is
more slowly varying.

The PY approximation, on the contrary, becomes less accurate when the
field's width increases, as is seen in Figure \ref{cap:nontinyreplusive}.
This is because the PY expansion about the bulk density and extrapolation
into the region where the external field is non-vanishing becomes less and
less justified when the range of the potential increases. This problem with
the PY approximation becomes much more acute for attractive potentials,
where its errors are magnified by the large Boltzmann factor in eq \ref
{eq:PY}, and the results are so poor that we do not show them in Figure \ref
{cap:nontinyattractive}. Indeed, the PY approximation for nonuniform fluid
is hardly ever applied in practice except for {}strongly repulsive
potentials, where the value of $\tilde{\rho}^{{\bf r}}$ in the repulsive
region is essentially irrelevant.

\begin{figure}[tbp]
\includegraphics[%
  width=0.95\columnwidth]{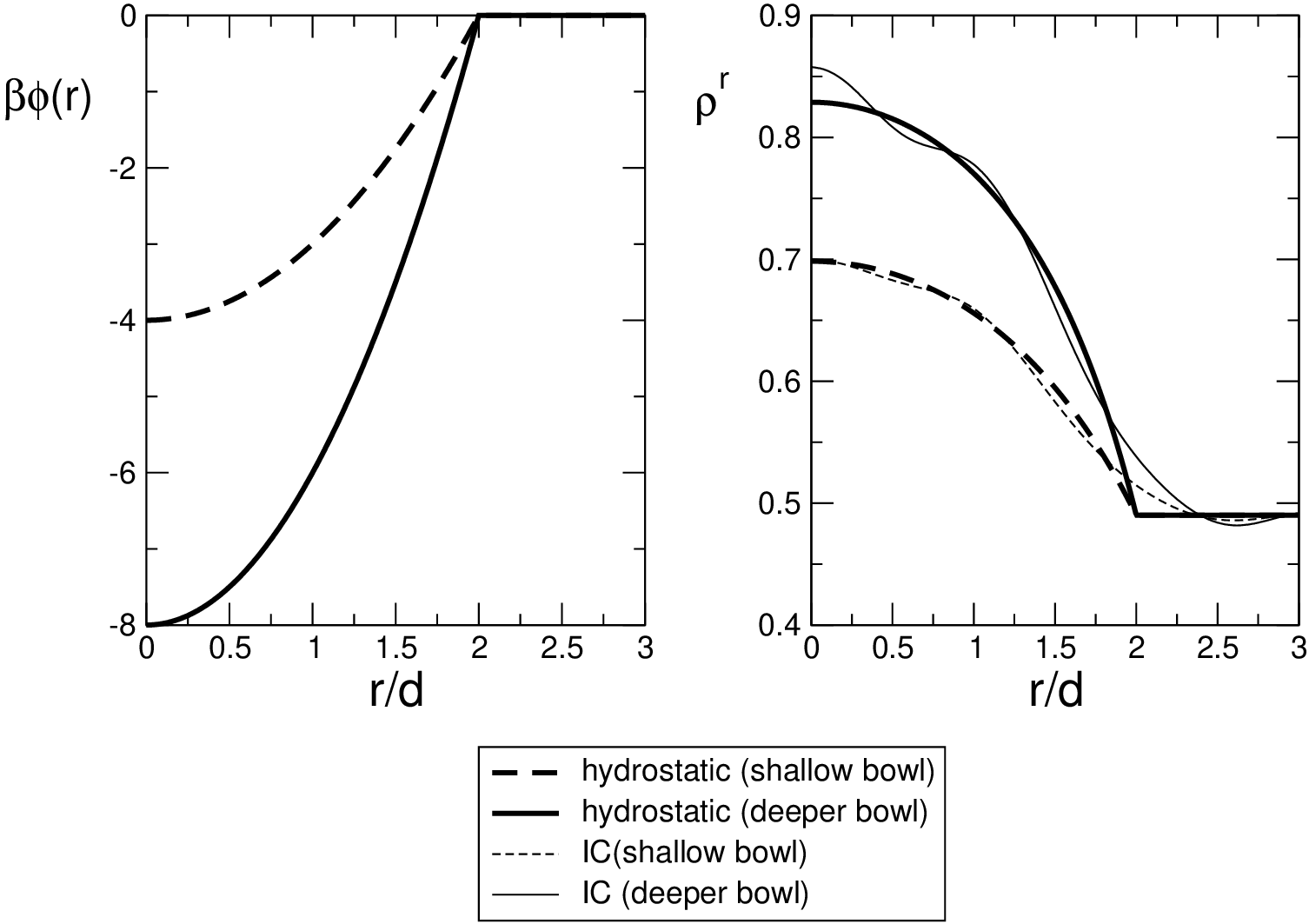}
  
\caption{The left graph gives the bowl potentials $\beta \phi (r)=A(r^{2}-4)$,
with $A=1$ for the shallow bowl, $A=2$ for the deeper bowl. The right
graph compares the hydrostatic densities $\rho ^{{\bf r}}$ of the external
fields to the $\tilde{\rho}^{{\bf r}}$ given by the IC equations.}
\label{cap:bowl-ref}
\end{figure}

The IC approximation is able again to correct the negative densities given
by HLR and, most notably, to capture the highly nontrivial density profile
{}inside both positive and negative step potentials due to the packing of
the hard spheres. However, for spherical cavities, although it is known that
the PY (and HLR) approximation consistently predicts a contact density lower
than the exact value, \cite{hansenmac,chen04} the IC noticeable
over-corrects the contact densities. This has a deleterious effect on the
rest of the profile, especially near the cavity region.

\begin{figure}[tbp]
\includegraphics[%
  width=0.90\columnwidth]{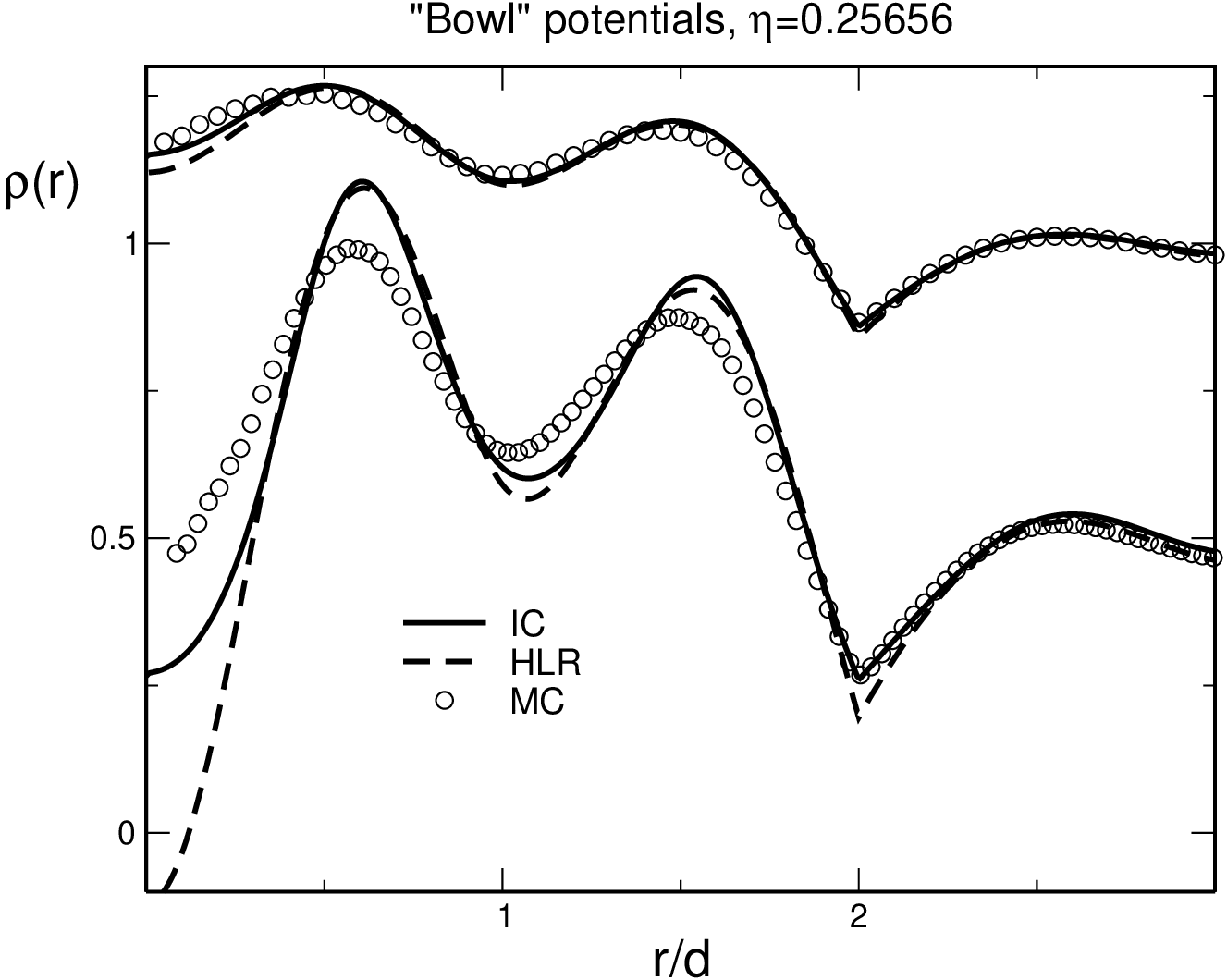}

\caption{Density responses to attractive parabolic bowl potentials. Here
$\eta =0.25656$.}
\label{cap:bowl-rho}
\end{figure}

In Figures \ref{cap:bowl-ref} and \ref{cap:bowl-rho} we consider the density
response of a hard sphere fluid with bulk packing fraction $\eta =0.25656$
to two deep {\em attractive parabolic bowl potentials.} Figure \ref
{cap:bowl-ref} shows the bowl potentials on the left and the reference
densities $\rho ^{{\bf r}}$ and $\tilde{\rho}^{{\bf r}}$ for the HLR and IC
equations on the right. Results for the HLR and IC approximations are
compared to MC simulations \cite{katsov01} in Figure \ref{cap:bowl-rho}. For
the shallower bowl potential, both the IC and HLR approximations agree well
with the Monte Carlo simulations. The reference density $\tilde{\rho}^{{\bf r%
}}$ of the IC approximation is very close to that of the hydrostatic density 
$\rho ^{{\bf r}}$, as can be seen in the right graph of Figure \ref
{cap:bowl-ref}, except that $\tilde{\rho}^{{\bf r}}$ varies smoothly near
the edge of the bowl, while the hydrostatic density has a discontinuous
derivative.

For the deeper bowl potential, both approximations deviate noticeably from
the simulation data, but nonetheless capture the nontrivial oscillatory
density profile inside the bowl. In particular, both reproduce the density 
{\em minimum} at the center of the bowl, where the external field is
actually most attractive, due to nonlocal effects from packing of the hard
spheres. However, the HLR density becomes negative at the bottom of the
bowl, while the IC density remains positive, though somewhat lower than the
MC result. The reference density $\tilde{\rho}^{{\bf r}}$ for the IC method
for the deeper bowl potential has more oscillations than that of the shallow
bowl potential, and exhibits a maximum at the center, which is the key for
keeping its predicted full density positive. Once outside the bowl, all
approximations agree well with the MC result.

\begin{figure}[tbp]
\includegraphics[%
  width=0.915\columnwidth]{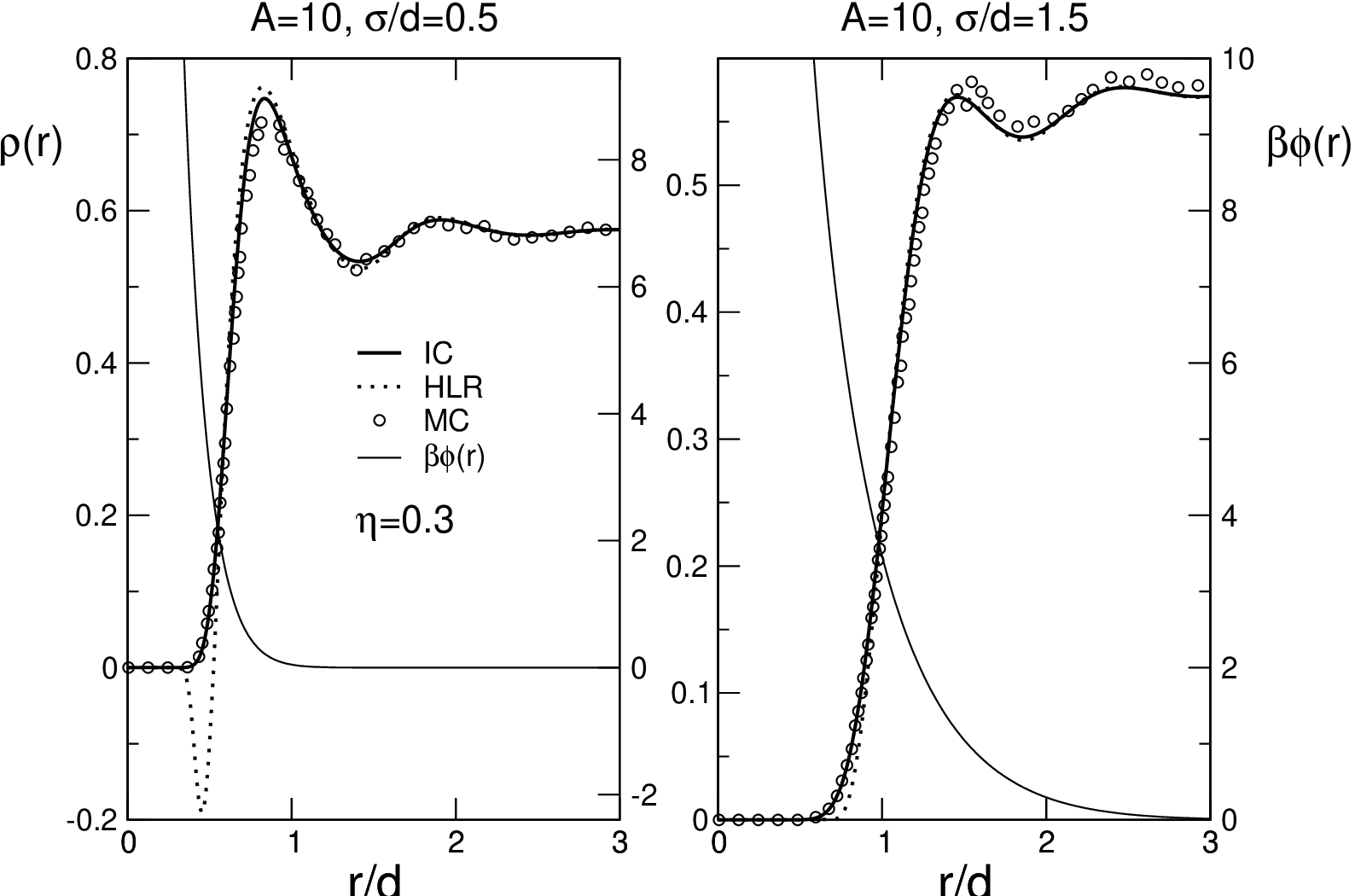}
  
\caption{The density response to two external fields of the form $A\text{erfc%
}(r/\sigma )/r$. Note that the left vertical axis refers to the density
curves while the right vertical axis labels $\beta \phi (r)$. Here $\eta =0.3$%
.}
\label{cap:erf}
\end{figure}

Figure \ref{cap:erf} plots the density response to a soft continuous
repulsive potential of the form $\beta \phi (r)=A\text{erfc}(r/\sigma )/r.$
This potential is important in our theory of ionic fluids, \cite
{ionicfluid04} but for our purposes here just serves as an example of a soft
repulsive potential. Here all the approximate results agree quite well with
the simulations, except that the HLR equation again shows a narrow negative
density region for the more rapidly varying potential (left graph of Figure 
\ref{cap:erf}).

\begin{figure}[tbp]
\includegraphics[%
  width=0.88\columnwidth]{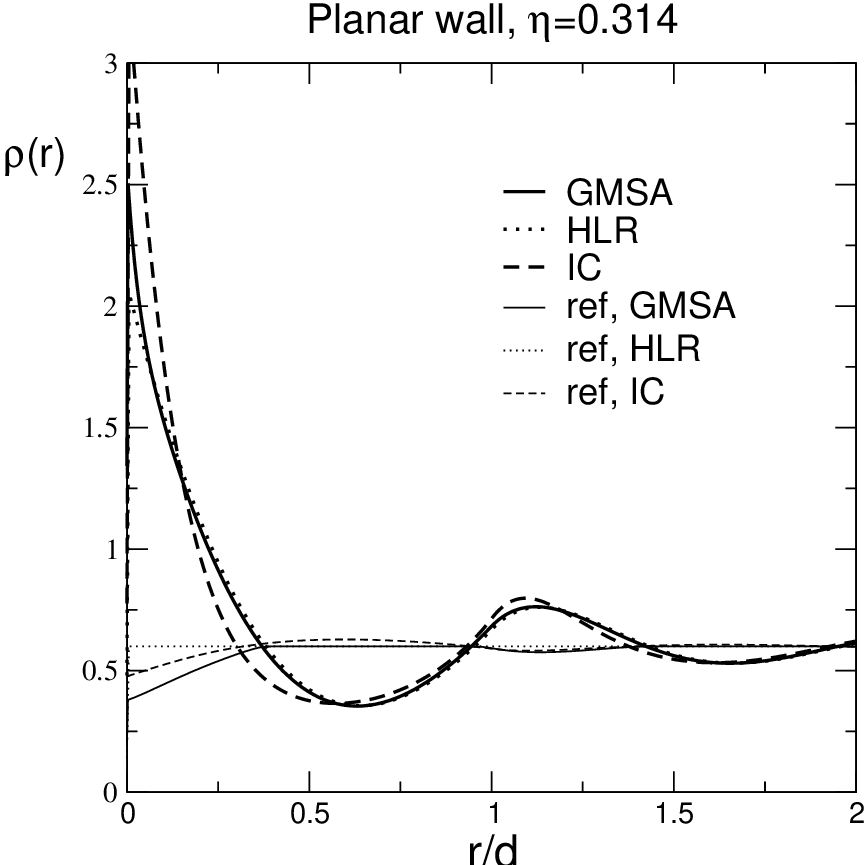}
  
\caption{Density response to one planar hard wall positioned at $z=0$. The
bulk packing fraction is $\eta =0.314$. The ``ref'' curves give the $\tilde{
\rho}^{z}$ used with the different approximations. The $\tilde{\rho}^{z}$
for the GMSA approximation is obtained by using the GMSA $\rho (z)$ in the IC
equation and solving inversely for the corresponding $\tilde{\rho}^{z}$.}
\label{cap:1wall}
\end{figure}

Finally, in Figure \ref{cap:1wall}, the density response to a planar hard
wall determined using the IC and HLR approximations is compared to the
results of the generalized mean spherical approximation (GMSA).\cite
{waismanMSA} The GMSA fits the contact density at the wall to the bulk
equation of state using an exact sum rule, and is known to be very accurate
for such systems. Thus it can be used as a benchmark for the other
approximations. As is well known, the HLR approximation (equivalent in this
case to the PY approximation) agrees well with the GMSA except for its
consistent underestimate of the contact density.

The IC result is now significantly worse than it was in the other cases. The
main problem arises from a severe overestimate of the contact density, which
goes on to spoil the rest of the density profile. The state shown is at a
moderate packing fraction $\eta =0.314$ and the errors get even larger at
higher densities. This problem is similar to that seen for the cavity
potentials presented before.

The IC method can also be used in an inverse way to determine what $\tilde{%
\rho}^{{\bf r}}$ is needed to obtain a given density $\rho ({\bf r})$ as a
solution. We determined the $\tilde{\rho}^{{\bf r}}$
associated with the accurate GMSA $\rho ({\bf r})$ in this way, and it can
be seen in Figure \ref{cap:1wall} that this GMSA $\tilde{%
\rho}^{{\bf r}}$ deviates from the bulk density (used in the
HLR equation) only very near the wall. The IC $\tilde{\rho}^{{\bf r}}$ shows a
similar deviation, but has more oscillations away from the wall and dips
less low near the wall.

This example shows that relatively small changes in $\tilde{\rho}^{{\bf r}}$
near the wall can have large effects on the predicted density profile near the
wall in the SLR equation. The fact that the HLR choice, clearly appropriate for
slowly varying fields, continues to give reasonably good results for single
hard walls and hard cavities seems somewhat fortuitous, as illustrated by
the errors HLR produces for rapidly varying but finite repulsive potentials.
See, e.g., the left graph in Figure \ref{cap:erf}. The IC
method, which gives very good results in most other limits, correctly
predicts a positive correction to the HLR/PY contact density but
overestimates its magnitude. Only a small change in $\tilde{\rho}^{{\bf r}}$
would be needed to produce very good results, but the IC prescription is not
able to determine this in advance.

\section{Final remarks}

\label{sec:Final remarks}The SLR equation provides a versatile framework for
computing the density response of hard sphere fluids to general external
fields. Since it satisfies the exact shifting property of the grand
ensemble, it can accurately describe two important limits: slowly varying
external fields (hydrostatic limit) and the opposite limit where field can
be very rapidly varying but only perturbs the fluid in tiny or narrow
regions. Errors in specific predictions arising from the linear truncation
in the SLR equation in other cases can be minimized by expanding about a
different uniform system at every point in space.

In practice there could be different prescriptions for how these local
uniform systems could best be chosen in particular applications and for
specific properties. In many cases the simple local HLR choice is quite
sufficient. However this has problems for rapidly varying but finite fields
and the SLR equation allows other choices. A general idea often used in
other expansions in liquid state theory is to choose a reference density
$\tilde{\rho}^{{\bf r}}$ that at least minimizes the quadratic correction to
the SLR equation in eq \ref{eq:Qexpand}. However, this is very complicated,
and there will still be unknown contributions from the higher order terms.

Here, as a first attempt, we have devised a computationally efficient
insensitivity criterion (IC), based on the idea that the SLR equation
should be insensitive to small variations of $\tilde{\rho}^{{\bf r}}$. This
property would be exactly satisfied if all terms in the expansion were taken
into account, and by imposing it self consistently on the SLR equation we
hope to generate a truncation where the contributions from higher order
terms are indeed small. The resulting IC method is extremely successful in
correcting the negative density regions that the HLR often exhibits for
rapidly varying finite fields, and although not exact, the IC method also
shows considerable improvement over HLR for the tiny field cases. Moreover,
the IC method is exact for narrow slits as the slit width $L_{s}\rightarrow
0 $, while the HLR and PY approximations have very significant errors in
this limit.

However, the IC method tends to overestimate the contact value of the
density response to single hard core cavities of all sizes, and this damages
the accuracy of the rest of the density profile. In practice this need not
be a significant limitation, since these cases are reasonably well treated
by the HLR and PY approximations. Other specific conditions for hard core
potentials, such as the sum rule used in the GMSA approximation, could be
taken into account to improve the IC method in this limit.

But it seems conceptually worthwhile to try to choose
$\tilde{\rho}^{{\bf r}}$ more generally within the SLR framework so that
accurate results naturally arise in this limit as well. To that end we
believe it would be profitable to further study the tiny field limit,
where similar problems are encountered, to gain additional insights into
the optimal choice of the reference density for the SLR equation. We
also need more information about the analytic nature and uniqueness of
solutions the IC or similar criteria can provide. We have solved the
resulting nonlinear equations numerically by iteration and have found a
stable self consistent solution. But there could be other solutions, or
other branches that only a small change in the IC could favor. We have
preliminary evidence that in the hard wall limit an alternative branch
may exist for high density states that could give much better results, and
plan further research along these lines.

\section{Acknowledgments}

This work was supported by the National Science
Foundation through Grant CHE-0111104. We are grateful to Michael
Fisher and Jim Henderson for helpful comments.

\appendix*
\section{Insensitivity condition for the SLR equation}

\label{sec:Insensitivity-condition}

After carrying out the functional derivative $\delta \rho ({\bf r}%
_{1})/\delta \tilde{\rho}^{{\bf r}_{2}}$ in eq \ref{eq:mhlrjdwdiff} on the
$\rho ({\bf r}_{1})$ given by the SLR equation \ref{eq:masterequation}, one
can rewrite the resulting expression as
\begin{widetext} 
\begin{equation}
\begin{array}[b]{l}
\medskip 
\displaystyle \int %
d{\bf r}_{3}\left[ \delta ({\bf r}_{1}-{\bf r}_{3})-\tilde{\rho}^{{\bf r}%
_{1}}e^{-\beta \tilde{\phi}^{{\bf r}_{1}}({\bf r}_{1})}c^{(2)}(r_{13};\tilde{%
\rho}^{{\bf r}_{1}})\right] 
{\displaystyle {\delta \rho ({\bf r}_{3}) \over \delta \tilde{\rho}^{{\bf r}_{2}}}}%
\\ 
=\medskip \tilde{\rho}^{{\bf r}_{1}}e^{-\beta \tilde{\phi}^{{\bf r}_{1}}(%
{\bf r}_{1})}\delta ({\bf r}_{1}-{\bf r}_{2})\left\{ \dot{c}^{(1)}(\tilde{%
\rho}^{{\bf r}_{1}})-%
\displaystyle \int %
\right. d{\bf r}_{3}c^{(2)}(r_{13};\tilde{\rho}^{{\bf r}_{1}}) \\ 
+\left. \medskip 
\displaystyle \int %
d{\bf r}_{3}\left[ \dot{c}^{(1)}(\tilde{\rho}^{{\bf r}_{1}})c^{(2)}(r_{13};%
\tilde{\rho}^{{\bf r}_{1}})+\dot{c}^{(2)}(r_{13};\tilde{\rho}^{{\bf r}%
_{1}})\right] (\rho ({\bf r}_{3})-\tilde{\rho}^{{\bf r}_{1}})\right\} 
\end{array}
\label{eq:mhlrjdwdiffdercoll}
\end{equation}

A special case of eq \ref{eq:Cndiff} relating the $n-1$ and $n$th order
direct correlation functions can be written as \cite{percus64} 
\begin{equation}
\dot{c}^{(n-1)}({\bf r}_{1},...,{\bf r}_{n-1};\rho )=\int d{\bf r}%
_{n}c^{(n)}({\bf r}_{1},...,{\bf r}_{n-1},{\bf r}_{n};\rho ).
\label{eq:cnn-1}
\end{equation}
Using this on the right hand side of eq \ref{eq:mhlrjdwdiffdercoll}, the
first two terms in the curly brackets cancel. Thus, the requirement that
$\delta \rho ({\bf r}_{3})/\delta \tilde{\rho}^{{\bf r}_{2}}$ in eq. \ref
{eq:mhlrjdwdiffdercoll} vanish for all ${\bf r}_{2}$ and ${\bf r}_{3}$ as in
eq \ref{eq:mhlrjdwdiff} then implies that 
\begin{equation}
\int d{\bf r}_{3}\left[ \dot{c}^{(1)}(\tilde{\rho}^{{\bf r}%
_{1}})c^{(2)}(r_{13};\tilde{\rho}^{{\bf r}_{1}})+\dot{c}^{(2)}(r_{13};\tilde{%
\rho}^{{\bf r}_{1}})\right] (\rho ({\bf r}_{3})-\tilde{\rho}^{{\bf r}%
_{1}})=0, \label{eq:wtermzero}
\end{equation}
for all ${\bf r}_{1},$ from which eq \ref{eq:mhlrjdwcriteriaave} follows.
Note that by using eq \ref{eq:cnn-1}, the equation above can be written as 
\begin{equation}
\displaystyle \int %
d{\bf r}_{2}%
\displaystyle \int %
d{\bf r}_{3}\left[ c^{(2)}(r_{12};\tilde{\rho}^{{\bf r}_{1}})c^{(2)}(r_{13};%
\tilde{\rho}^{{\bf r}_{1}})+c^{(3)}({\bf r}_{1},{\bf r}_{2},{\bf r}_{3};%
\tilde{\rho}^{{\bf r}_{1}})\right] \,(\rho ({\bf r}_{3})-\tilde{\rho}^{{\bf r%
}_{1}})=0  \label{eq:wtermint}
\end{equation}

This could also be derived by making the following approximation for $Q({\bf %
r})$ in eq \ref{eq:Qexpand}: 
\begin{equation}
Q({\bf r}_{1})\approx (\rho ({\bf r}_{1})-\tilde{\rho}^{{\bf r}_{1}})\int d%
{\bf r}_{2}\int d{\bf r}_{3}\left[ c^{(2)}(r_{12};\tilde{\rho}^{{\bf r}%
_{1}})c^{(2)}(r_{13};\tilde{\rho}^{{\bf r}_{1}})+c^{(3)}({\bf r}_{1},{\bf r}%
_{2},{\bf r}_{3};\tilde{\rho}^{{\bf r}_{1}})\right] (\rho ({\bf r}_{3})-%
\tilde{\rho}^{{\bf r}_{1}}),
\end{equation}
\end{widetext} 
i.e., by assuming that $\rho ({\bf r}_{2})$ differs little from $\rho ({\bf r%
}_{1})$ in the region of integration near ${\bf r}_{1}$ in the definition
for $Q$. Requiring that this approximation for the quadratic term vanish
then gives eq \ref{eq:wtermint}.


\begin{thebibliography}{99}
\bibitem{percusnote}  See eq 5.18 in Reference \onlinecite{percus64}, p. II-66.
Note there is a misprint in the sign of the last term of this equation.
\end{thebibliography}
\end{document}